\documentclass[twocolumn]{aastex631}

\usepackage{graphicx}	
\usepackage{amsmath}	
\usepackage{amssymb}	
\usepackage{xcolor}
\usepackage{multirow}
\usepackage{soul}

\def\omA{\omega_{\rm A}}

\def\bB{\boldsymbol{B}}

\def\kTI{k_{\rm TI}}

\def\Jweb{J_{\rm web}}
\def\Iweb{I_{\rm web}}
\def\Rweb{R_{\rm web}}
\def\Omweb{\Omega_{\rm web}}
\def\mt{m_{\rm t}}
\def\mb{m_{\rm b}}
\def\mHe{m_{\rm He}}

\def\tpump{t_0}
\def\bphipert{\overline{b}_{\phi}}
\def\bperttilde{\tilde{b}_{\phi}}
\def\Eq{Equation}

\def\beq{\begin{equation}}
\def\eeq{\end{equation}}

\begin{document}

\title{ Magnetic Webs in Stellar Radiative Zones}

\author{Valentin A. Skoutnev}
\affiliation{Physics Department and Columbia Astrophysics Laboratory, Columbia University, 538 West 120th Street New York, NY 10027,USA}
\affiliation{Max Planck Institute for Astrophysics, Karl-Schwarzschild-Str. 1, D-85741, Garching, Germany}

\author{Andrei M. Beloborodov}
\affiliation{Physics Department and Columbia Astrophysics Laboratory, Columbia University, 538 West 120th Street New York, NY 10027,USA}
\affiliation{Max Planck Institute for Astrophysics, Karl-Schwarzschild-Str. 1, D-85741, Garching, Germany}



\begin{abstract}

Rotational evolution of stellar radiative zones is an old puzzle. We argue that angular momentum transport by turbulent processes induced by differential rotation is insufficient, and propose that a key role is played by ``magnetic webs." We define magnetic webs as stable magnetic configurations that enforce corotation of their coupled mass shells, and discuss their resistance to differential torques that occur in stars. Magnetic webs are naturally expected in parts of radiative zones that were formerly convective, retaining memory of extinguished dynamos. For instance, red giants with moderate masses $M\gtrsim 1.3M_\odot$ likely contain a magnetic web deposited on the main sequence during the retreat of the central convective zone. The web couples the helium core to the hydrogen envelope of the evolving red giant and thus reduces spin-up of the contracting core. The magnetic field and the resulting slower rotation of the core are both consistent with asteroseismic observations, as we illustrate with a stellar evolution model with mass $1.6M_\odot$. Evolved massive stars host more complicated patterns of convective zones that may leave behind many webs, transporting angular momentum towards the surface. Efficient web formation likely results in most massive stars dying with magnetized and slowly rotating cores.

\end{abstract}



\section{Introduction}

Rotation plays a significant role in the lives of stars. Rotational gradients drive instabilities and mixing, which affect stellar evolution and chemical abundance patterns. At the end of a massive star's life, rotation of its core affects the final collapse and supernova explosion, as well as the angular momentum (AM) of the compact remnant. Stars develop differential rotation for a few reasons, including AM losses from the surface, interactions in binary systems, and changes in stellar structure (e.g. spin up of a contracting core and spin down of an expanding envelope). The radial profile of the angular velocity $\Omega(R)$ is governed by these processes together with AM transport across the star, which is poorly understood.

Asteroseismic observations provide constraints on the AM transport in evolving low-mass stars. Without AM transport, the (stably stratified) helium core of a red giant would rotate several orders of magnitude faster than its hydrogen envelope (which includes a radiative ``mantle" and an outer convective zone). Observations show only an order of magnitude faster rotation, indicating that most of the core AM is lost \citep{beck2012fast,mosser2012spin,deheuvels2012seismic,deheuvels2015seismic,di2018rotational,gehan2018core,tayar2019core,kuszlewicz2023mixed,hatt2024asteroseismic,li2024asteroseismic,mosser2024locked}. How the core sheds its AM is not settled. Transport enabled by hydrodynamical instabilities and waves is found to be inefficient \citep{eggenberger2012angular,ceillier2013understanding,marques2013seismic,cantiello2014angular,fuller2014angular}, implicating magnetohydrodynamic processes (for a review, see \citealt{aerts2019angular}).

Turbulent transport driven by magnetohydrodynamic instabilities in differentially rotating regions is most commonly invoked. One challenge is to find instabilities that are not inhibited by the strong compositional gradients surrounding the helium core \citep{spruit1999differential,heger2000presupernova,wheeler2015role}. A leading candidate has been the Tayler instability of toroidal fields generated by differential rotation \citep{tayler1973adiabatic,spruit1999differential,spruit2002dynamo,fuller2019slowing}. However, a recent revision of the linear stability analysis shows that the Tayler instability is suppressed in a shell surrounding the core \citep{Skoutnev_2024a,Skoutnev_2024b}. Furthermore, instability requires the radial magnetic field to be sufficiently weak $B_R\lesssim 3$\,G. Much stronger fields up to $B_R\sim 10^5$\,G may be left behind by a core-convection dynamo during the MS and compressed up to $B_R\sim 10^6-10^7$\,G during the red giant branch (RGB) \citep{cantiello2016asteroseismic}, possibly surviving into the white dwarf phase \citep{kissin2015spin,bagnulo2022multiple,camisassa2024main}. Asteroseismic observations indicate that strong remnant fields are indeed possible, with $B_R\gtrsim 10^5$\,G inferred from dipole-mode suppression \citep{fuller2015asteroseismology,stello2016prevalence} and $B_R\gtrsim 40$\,kG from mode-splitting \citep{li2022magnetic,deheuvels2023strong, li2023internal}. These fields far exceed the typical $B_R\sim 10^{-2}$\,G predicted by AM transport models based on the Tayler instability \citep{fuller2019slowing} and can easily quench turbulent transport.

On the other hand, strong magnetic fields threading the entire star would be extremely efficient at transporting AM. In particular, axisymmetric magnetic fields are known to maintain corotation of coupled mass shells when their Alfv\'enic timescale is shorter than the timescale to pump differential rotation \citep{mestel1987magnetic,spruit1998birth,maeder2014magnetic,kissin2015rotation,kissin2018rotation,takahashi2021modeling,gouhier2022angular}. This implies that global fields as weak as $B_R\sim10^{-4}\;$G can force entire radiative zones of evolved low-mass stars to corotate. However, such global coupling is inconsistent with observations of red giants, which require significant differential rotation to occur somewhere between the core and the envelope \citep{di2018rotational,klion2017diagnostic,fellay2021asteroseismology}.

\begin{figure}
    \centering
     \includegraphics[width=\linewidth]{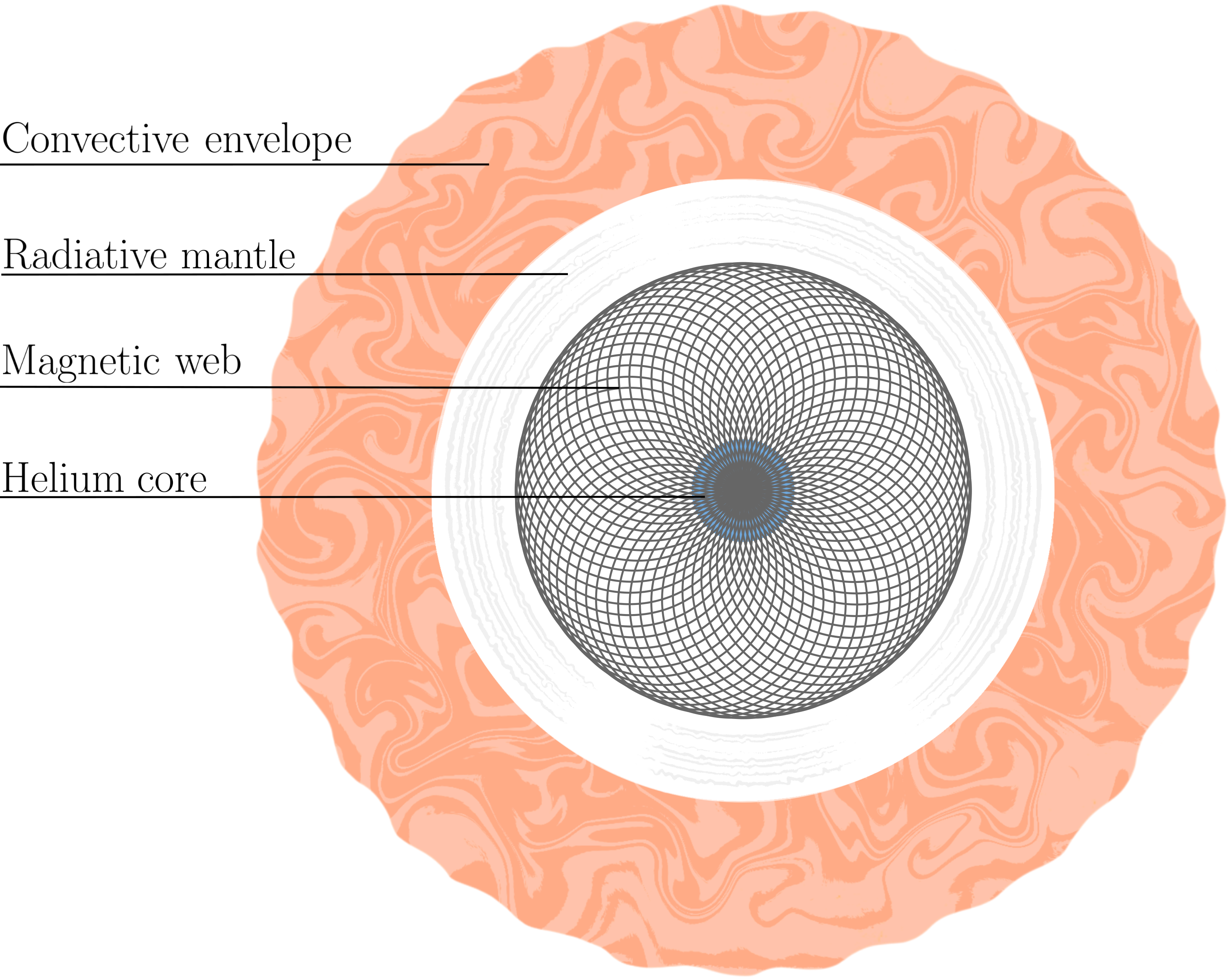}
    \caption{
    Schematic picture of a red giant hosting a magnetic web deposited during MS evolution. The magnetic web encloses the helium core and the lower radiative mantle, enforcing their corotation. The web-free, upper radiative mantle is free to rotate differentially, with a profile of $\Omega(R)$ regulated by turbulent transport. Surrounding the mantle is a large, slowly rotating, convective envelope.
    }
    \label{fig:MagneticWeb}
\end{figure}

This Letter proposes that the puzzle of AM transport is resolved if stellar radiative zones include two types of regions:
(1) regions with solid-body rotation enforced by a stable and sturdy magnetic configuration, which we term a ``magnetic web," and 
(2) differentially rotating regions where AM is exchanged via turbulent transport. 
The rotational evolution of stars within this framework depends on the history of magnetic web formation by dynamo activity in past phases of stellar evolution. In isolated stars, magnetic webs are relicts of receding convective zones. For instance, in stars with $M\gtrsim 1.3 M_\odot$, magnetic fields are deposited by the retreating core convection zone during the main sequence (MS) phase, so at later evolutionary stages a web covers the previously convective central region (Figure~\ref{fig:MagneticWeb}). Stars interacting with external bodies (in binaries and their mergers, or stars engulfing planets) can have particularly rich histories of dynamo episodes and magnetic web formation.

\newpage

\section{Magnetic fields and corotation}

\label{sec:BasicWebs}

Magnetic fields deposited in a radiative zone are generally thought to relax into stable configurations (for a review, see \citealt{braithwaite2017magnetic}). Once formed, stable magnetic configurations can persist on evolutionary timescales without being affected by magnetic diffusion \citep{cantiello2016asteroseismic}. The primary threat to their survival is differential rotation, which would deform configurations, make them unstable, and eventually erase them through rotational smoothing (e.g. \citealt{radler1986effect,wei2015obliquely}). Sufficiently strong fields can form sturdy magnetic configurations that inhibit the development of differential rotation and only weakly deform from their equilibria. Below we evaluate the minimum magnetic fields that meet this requirement and estimate their values in MS and evolved stars.

\subsection{A toy model}
\label{toy}

\begin{figure}
    \centering
    \includegraphics[width=0.9\linewidth]{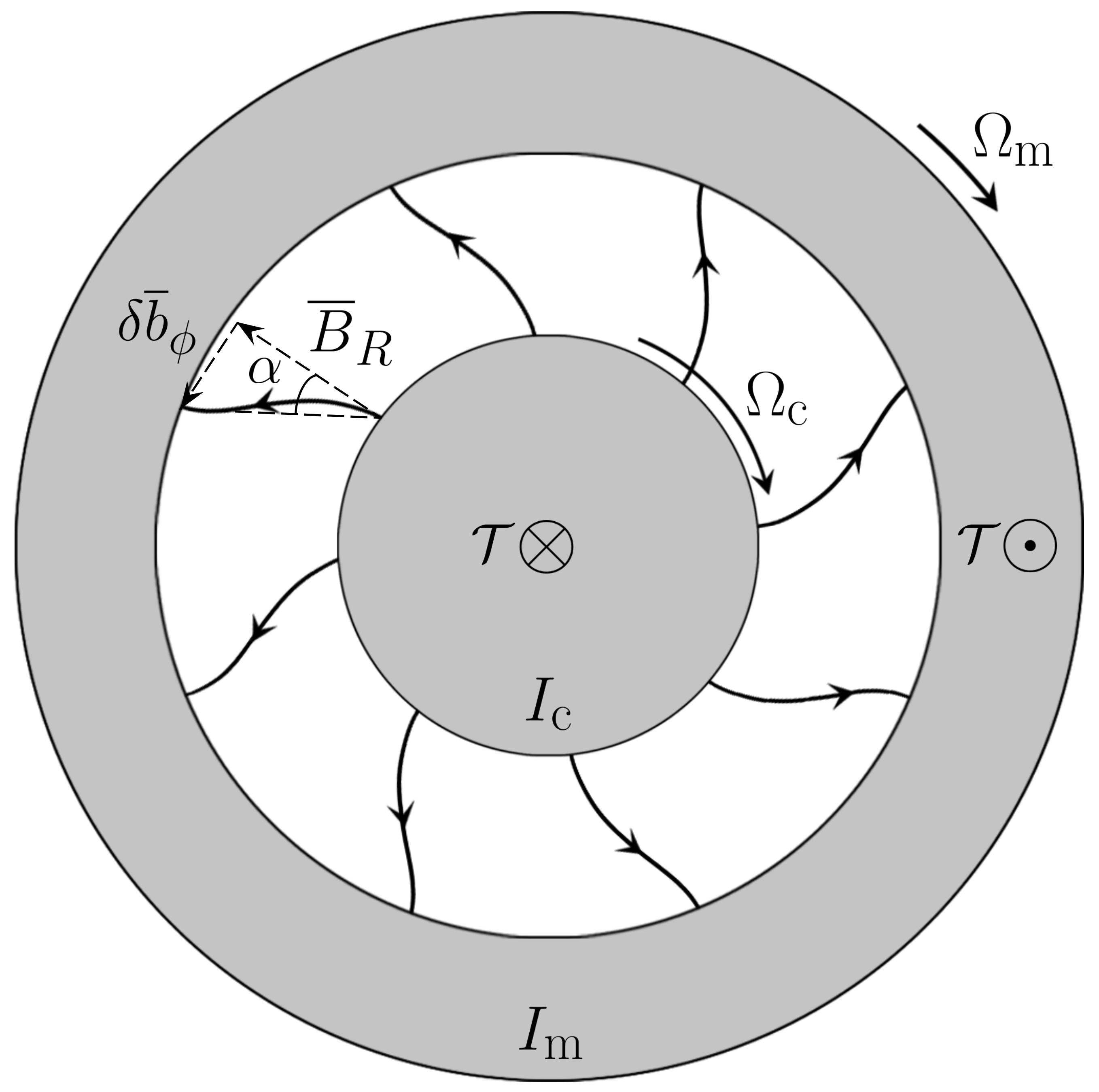}
    \caption{
    Toy model of an evolving core and mantle connected by a radial magnetic field.
    }
    \label{fig:TZModel}
\end{figure}

The basic response of a stable magnetic configuration to the pumping of differential rotation can be understood through a toy model. Consider a constant, axisymmetric, radial\footnote{The model will be extended below to include a background toroidal field $\overline{B}_\phi$.} magnetic field $\overline{B}_R$ connecting two rigidly rotating shells (the ``core'' and the ``mantle'')  as shown in Figure~\ref{fig:TZModel}. The shells have moments of inertia $I_{\rm c}$ and $I_{\rm m}$, rotation rates $\Omega_{\rm c}$ and $\Omega_{\rm m}$, and are initially corotating, $\Omega_{\rm c}=\Omega_{\rm m}$. The source pumping differential rotation will be modeled as torques $\pm{\cal T}$ acting on the inner/outer shells at time $t>0$. 

Any build up of differential rotation generates an axisymmetric toroidal field $\bphipert(t)$ and a restoring magnetic torque $R^3\overline{B}_R \overline{b}_\phi$ (ignoring geometrical factors), where $R$ is the core radius. The equations for the AM of the shells and $\bphipert$ (from magnetic induction) are: 
\begin{align}
    I_{\rm c} \dot\Omega_{\rm c}&=R^3\overline{B}_R  \bphipert+
    {\cal T},
    \\
    I_m \dot\Omega_{\rm m}&=-R^3\overline{B}_R  \bphipert-
    {\cal T},
    \\ 
    \dot{\overline{b}}_\phi&=\overline{B}_R(\Omega_{\rm m}-\Omega_{\rm c}), 
\end{align}
where the dot signifies a time derivative. 
These equations reduce to the driven oscillator equation for the deformation angle of the field $\alpha\equiv\bphipert/\overline{B}_R$:
\begin{equation}
 \ddot\alpha + \omA^2\alpha = 
 -\frac{{\cal T}}{I_{\rm eff}}, \qquad \omega_{\rm A}^2=\frac{R^3\overline{B}_R^2}{I_{\rm eff}},
\end{equation}
where $I_{\rm eff}^{-1}=I_{\rm c}^{-1}+I_{\rm m}^{-1}$. 

Differential rotation in isolated stars is typically pumped on evolutionary timescales. In the toy model shown in Figure~\ref{fig:TZModel}, this may be described as gradually appearing torques $\pm{\cal T}(t)$ that deposit AM $\sim \pm I_{\rm eff}\Omega_{\rm c}$ on a timescale $\tpump$: ${\cal T}(t)=I_{\rm eff}\Omega_{\rm c} \, t/\tpump^2$. Then, one finds
\beq
  \alpha(t)=\alpha_0\left[\frac{t}{\tpump}-\frac{\sin(\omega_{\rm A}t)}{\omega_{\rm A}\tpump}\right], \qquad \alpha_0=-\frac{\Omega_{\rm c}}{\omA^2\tpump}.
\eeq
For magnetic fields of interest, the Alfv\'en timescale $\omA^{-1}$ is much shorter than the pumping timescale, $\omA \tpump\gg 1$. Then, the oscillating term $\propto \sin(\omA t)$ quickly (at $t>\omA^{-1}$) becomes small compared to the smoothly growing term $\propto t/t_0$, which describes a quasistatic deformation in response to the slowly varying ${\cal T}$: 
\beq
\label{eq:toy_alpha_quasi}
    \alpha(t)\approx -\frac{{\cal T}(t)}{I_{\rm eff}\omA^2}.
\eeq

The condition $\omA \tpump\gg 1$ implies $|\dot\alpha|=|\Omega_{\rm c}-\Omega_{\rm m}|\ll\Omega_{\rm c}$, so the magnetic coupling prevents differential rotation. A more demanding condition for a sturdy magnetic configuration is a small deformation of the magnetic field, $|\alpha|\ll 1$. It is satisfied if $\omA^2\gg \Omega_{\rm c}/\tpump$, which requires 
\begin{align}
\label{eq:weakdeflections}
    \overline{B}_R\gg\sqrt{\frac{I_{\rm eff}\Omega_{\rm c}}{ R^3 \tpump}}.
\end{align}

This toy model illustrates how a magnetic web enforces corotation in stellar interiors. Realistic stable magnetic configurations have a toroidal component (necessary for their stability) and their sturdiness condition slightly differs from that in \Eq~(\ref{eq:weakdeflections}): now quasistatic deformations $\bphipert$ must remain small compared to the total background field $\overline{B}$. Since stable configurations have $\overline{B}_\phi\gtrsim \overline{B}_R$, the required radial field becomes
\beq
  \label{eq:weakdeflections1}
    \overline{B}_R \gg \overline{B}_{\rm web}
    =\sqrt{\frac{I_{\rm eff}\Omega_{\rm c}\overline{B}_R}{ R^3 \tpump \overline{B}_\phi}}.
\eeq
It is also possible that configurations tolerate large deformations, potentially until the total toroidal field $\bphipert+\overline{B}_\phi$ reaches the Tayler instability threshold $\sim\overline{B}_R(\kTI R)$ \citep{braithwaite2009axisymmetric}, where $\kTI$ is the lowest unstable radial wavenumber. Such durability would imply that configurations with radial fields even weaker than $\overline{B}_{\rm web}$ in \Eq~(\ref{eq:weakdeflections1}) may be able to act as magnetic webs.

In a star with a continuum of magnetically coupled mass shells $\mb\leq m\leq \mt$ with a characteristic radius $\Rweb$, one can estimate $\overline{B}_{\rm web}$ in \Eq~(\ref{eq:weakdeflections1}) as
\begin{align}
\label{eq:Bweb}
    \overline{B}_{\rm web}
    \sim
    \sqrt{\frac{\Jweb}{\Rweb^3 \tpump}\frac{\overline{B}_R}{\overline{B}_\phi}},\quad\Jweb=\int_{\mb}^{\mt}j(m)dm,
\end{align}
where $j(m)$ is the specific AM. Rotation of the web-covered region $\Omweb(t)=\Jweb(t)/\Iweb(t)$ is determined by its AM $\Jweb$ and moment of inertia $\Iweb$. 

\subsection{Full three-dimensional magnetic response}
\label{sec:Responce}

We now discuss the 3D magnetic response and how it effectively reduces to the toy model described above. Consider a stable magnetic configuration contained in a stably stratified fluid initially rotating as a solid body with rate $\Omega$. In general, the magnetic configuration with total field $\bB$ has both axisymmetric $\overline{\bB}$ and non-axisymmetric $\tilde{\bB}$ components. Suppose now that differential rotation is pumped by an axisymmetric torque with density $\tau$. The response of the magnetized fluid is convenient to view separately for $\overline{\bB}$ and $\tilde{\bB}$.

We are interested in the typical regime of fast rotation in stars, $\Omega\gg\overline{\omega}_{\rm A},\tilde{\omega}_{\rm A}$, where $\overline{\omega}_{\rm A}=\overline{B}_R/\sqrt{4\pi\rho}R$ and $\tilde \omega_{\rm A}=\tilde B_R/\sqrt{4\pi\rho}R$. Then, $\overline{\bB}$ and $\tilde{\bB}$ respond to perturbations on different timescales (for a review, see \citealt{jault2015waves}). A magnetic configuration $\bB=\tilde{\bB}$ with a vanishing average over $\phi$ responds on the timescale $t_{\rm ms}\sim \Omega/\tilde \omega_{\rm A}^2$, which describes ``magnetostrophic" motions with balanced Coriolis and Lorentz forces. The Coriolis force tends to arrest motions perpendicular to $\boldsymbol{\Omega}$ and slows down magnetic waves by the factor $\tilde{\omega}_{\rm A}/\Omega$ compared to usual Alfv\'en waves. By contrast, an axisymmetric configuration $\bB=\overline{\bB}$ responds on the Aflv\'enic timescale $t_{\rm A}\sim 1/\overline{\omega}_{\rm A}$. This is the timescale for a special class of motions known as toroidial Aflv\'en waves that bypass the inhibitory effects of Coriolis forces.\footnote{This can also be seen from the analysis of linear perturbations  with large radial wavenumbers $k\gg1/R$ in an axisymmetric field. Then, using expansion in spherical harmonics $Y_{lm}$, one finds the perturbation frequency $\omega$ from the dispersion relation $k^2R^2\overline{\omega}_{\rm A}^2=\omega^2-2m\Omega\omega/l(l+1)$ \citep{levin2004hydromagnetic}. The response is Alfv\`enic $\omega\propto\overline{\omega}_{\rm A}$ for axisymmetric perturbations $m=0$.} 

In the presence of both $\overline{\bB}$ and $\tilde{\bB}$, the response is dominated by $\overline{\bB}$ ($t_{\rm A}<t_{\rm ms}$) if 
\begin{align}
    \frac{\overline{B}_R}{\tilde B_R}> \frac{\tilde{\omega}_{\rm A}}{\Omega},
\end{align}
which is likely satisfied since $\tilde\omega_{\rm A}/\Omega\ll 1$. The response of $\overline{\bB}$ is discussed in Appendix~\ref{axisym_response}. It involves small-amplitude torsional Alfv\'en waves and is dominated by a quasistatic deformation of the magnetic configuration,
\beq
\label{eq:DeltaB}
    |\Delta \bphipert|\sim \frac{m_{\rm web}\Omega}{\Rweb \overline{B}_R\tpump}.
\eeq
The deformation is effectively the same as in the toy model (\Eq~\ref{eq:toy_alpha_quasi}), and hence the minimum radial field $\overline{B}_{\rm web}$ needed for a web is similar to that of Equation~(\ref{eq:Bweb}).

The conditions for a magnetic web are changed for configurations with a very weak axisymmetric component, $\overline{B}_R/\tilde{B}_R<\tilde{\omega}_{\rm A}/\Omega$, whose coupling timescale is $t_{\rm ms}$. The response of $\bB=\tilde{\bB}$ to a torque applied on a timescale $\tpump$ is the superposition of the quasistatic deformation given by \Eq~(\ref{eq:DeltaB}) and transient magnetostrophic waves with amplitudes $\delta \Omega\sim \Omega(t_{\rm ms}/\tpump)^2$ and $\delta \bperttilde\sim \tilde B_R\Omega t_{\rm ms}(t_{\rm ms}/\tpump)^2$. The corotation condition $\delta\Omega\ll \Omega$ is satisfied when $t_{\rm ms}\ll \tpump$. The more stringent condition for weak deformations is $\Delta \bperttilde+\delta \bperttilde\ll \tilde{B}_\phi$. Unlike axisymmetric configurations, the wave component $\delta \bperttilde$ is now large because the response occurs on the longer timescale $t_{\rm ms}\gg \tilde{\omega}_{\rm A}^{-1}$. The condition for a sturdy magnetic configuration then becomes
\beq
  \frac{m_{\rm web}\Omega}{\Rweb\tilde{B}_R\tpump} + \tilde{B}_R \frac{\Omega\, t_{\rm ms}^3}{\tpump^2} \ll \tilde{B}_\phi,
\eeq
which one can rewrite as
\beq
  \Omega \tpump +\frac{\Omega^4}{\tilde{\omega}_{\rm A}^4}\ll \frac{\tilde B_\phi}{\tilde B_R} (\tilde{\omega}_{\rm A}\tpump)^2.
\eeq
It defines a lower limit $\tilde{\omega}_{\rm A,min}\sim (\tilde B_R/\tilde B_\phi)^{1/6}\Omega^{2/3}\tpump^{-1/3}$ which satisfies $\tilde{\omega}_{\rm A,min}^4 \ll \Omega^3/\tpump$ (corresponding to the wave-dominated regime of the response, $\delta \bperttilde\gg \Delta \bperttilde$). The condition $\tilde{\omega}_{\rm A}\gg\tilde{\omega}_{\rm A,min}$ requires $\tilde{B}_R\gg\tilde{B}_{\rm web}$ where
\begin{align}
\label{eq:weakdeflections2}
    \tilde B_{\rm web}\sim \sqrt{\frac{\Jweb}{ \Rweb^3\tpump }}\left(\Omega \tpump\frac{\tilde B_R}{\tilde B_\phi}\right)^{1/6}.
\end{align}

In summary, magnetic configurations with $\overline{B}_R>\overline{B}_{\rm web}$ or $\tilde{B}_R>\tilde{B}_{\rm web}$ are weakly deformed during stellar evolution; they enforce corotation. Magnetic configurations that satisfy both $\overline{B}_R<\overline{B}_{\rm web}$ and $\tilde{B}_R<\tilde{B}_{\rm web}$ may be destroyed by the pumped differential rotation.

\subsection{Webs in main sequence and evolved stars}
Differential torques naturally accompany contraction/expansion of mass shells in a star. Changes in stellar structure typically occur on evolutionary timescales, with a major contraction/expansion phase occurring on the RGB. In red giants of moderate mass ($1.3M_\odot\lesssim M\lesssim 2M_\odot$), magnetic fields left over from the MS core convection can cover both the helium core and a portion of the surrounding radiative mantle (Figure~\ref{fig:MagneticWeb}). For typical parameters, these regions remain locked in corotation if the radial field exceeds
\begin{align}
\nonumber
    \overline{B}_{\rm web}&\sim 30\;\mathrm{G}\left(\frac{J_{\rm web}}{10^{48}\rm \,erg\cdot s}\right)^{1/2}\left(\frac{\Rweb}{0.1R_{\odot}}\right)^{-3/2}\\
    & \times
    \left(\frac{\tpump}{10^8\,\rm yr}\right)^{-1/2}\left(\frac{\overline{B}_R}{\overline{B}_\phi}\right)^{1/2}.
\label{eq:Bweb1}
\end{align}
It is far below the  $\sim10^5$\,G fields at the edge of helium cores inferred from asteroseismology. Such strong fields make their coupled regions behave very much like a solid body, responding to torques with a tiny deformation by an angle $\alpha\sim \overline{B}_{\rm web}^2/\overline{B}_R^2\sim 10^{-7}$\,rad.

Massive stars experience significant expansion (by a factor of up to $\sim 3$) even earlier, when they are still on the MS \citep{ekstrom2008evolution,brott2011rotating,lander2012there}. Their radiative envelopes do not develop differential rotation if they contain a fossil field with a sufficiently strong $\overline B_R>\overline{B}_{\mathrm {web}}$. For example, consider a star with $M\sim 10M_\odot$, $\tpump\sim 10^7$\,yr, $R_{\rm web}\sim 5 R_\odot$, and $J_{\rm web}\sim 10^{52}$\,erg\,$\cdot$\,s. This gives $\overline{B}_{\mathrm {web}}\sim 30$\,G, smaller than the $\sim 1\,$kG surface fields observed in samples of magnetic massive stars \citep{wade2016mimes}.

Stellar winds can also be a significant driver of differential rotation, particularly in MS stars \citep{kudritzki2000winds,vink2022theory}. The surface spin-down torque  $\mathcal{T}=\dot M R_{\rm A}^2\Omega$ (e.g. \cite{weber1967angular}) due to mass loss with rate $\dot M$ pumps differential rotation on a timescale $t_0\sim MR^2/\mathcal{T}\Omega=(M/\dot M)(R/R_{\rm A})^2$, where $R_{\rm A}$ is the Alfv\'en radius. For Sun-like stars ($M\lesssim 1.3M_\odot$), this torque is transmitted through the convective envelope to the radiative core. The core will host a magnetic web if the radial component of its fossil field exceeds
\begin{align}
\nonumber
    \overline{B}_{\rm web}&\sim0.03\;\mathrm{G}\left(\frac{J_{\rm web}}{10^{48}\rm \,erg\cdot s}\right)^{1/2}\left(\frac{\Rweb}{0.5R_{\odot}}\right)^{-3/2}\\
    & \times
    \left(\frac{M/\dot M}{10^{14}\,\rm yr}\right)^{-1/2}\left(\frac{R_{\rm A}/R}{10}\right)\left(\frac{\overline{B}_R}{\overline{B}_\phi}\right)^{1/2}.
\label{eq:Bweb_winds}
\end{align}
A magnetic web in the Sun can explain helioseismic measurements of a nearly uniformly rotating radiative interior \citep{howe2009solar}, with the tachocline acting as a boundary layer that communicates torques between the convection zone and the web \citep{gough1998inevitability}.

In more massive stars ($M>1.3M_\odot$), the wind torque is applied to the radiative envelope. Their faster mass losses imply a shorter $t_0$ and a larger $\overline{B}_{\rm web}$, up to $\sim 30$\,G. A massive star with a web in its radiative envelope would effectively rotate as a solid body from the Alfv\'en radius $R_{\rm A}$ to the bottom of the envelope.

\section{A stellar model}
\label{sec:StellarModel}

\begin{figure}[h]
    \centering
    \includegraphics[width=\linewidth]{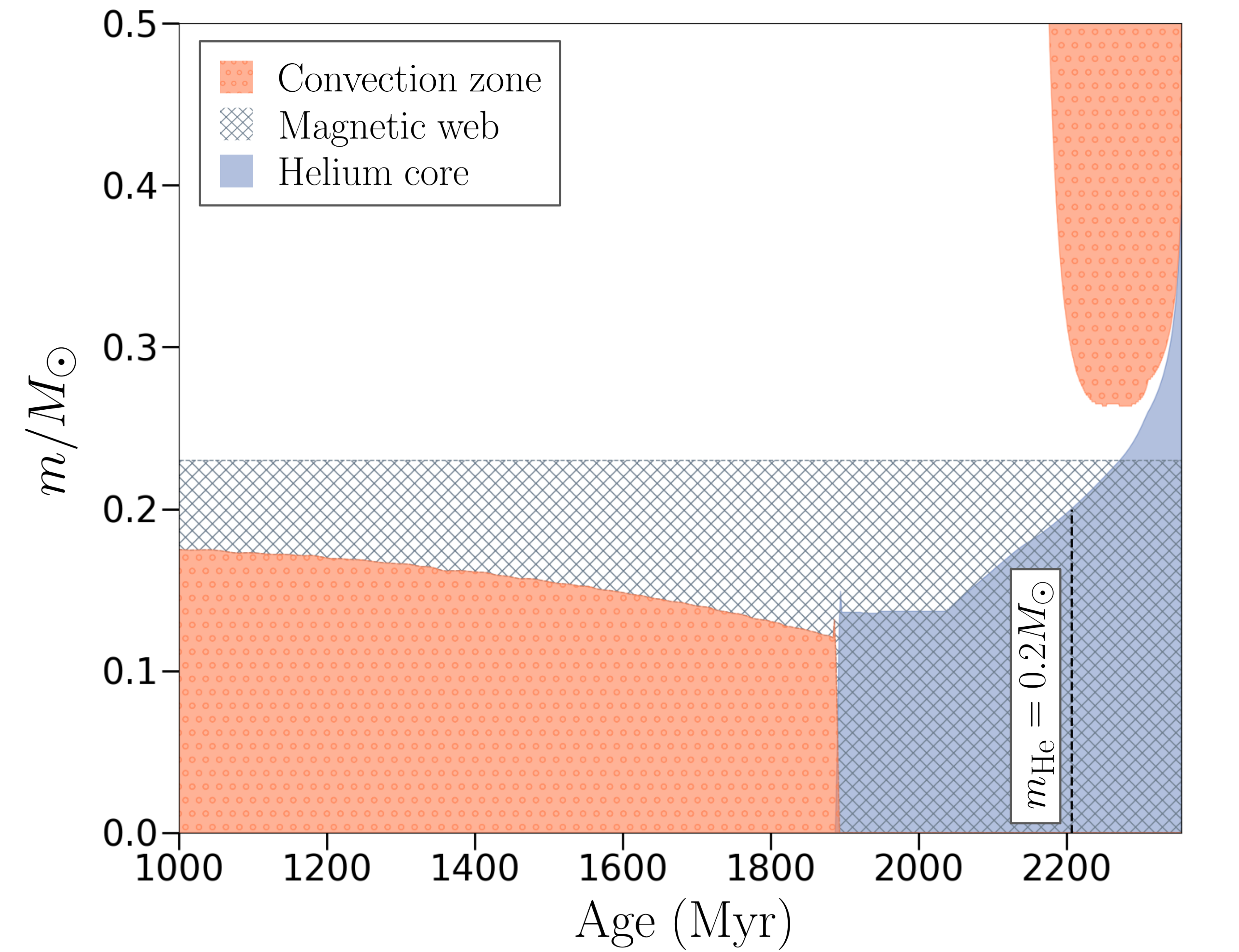}
    \includegraphics[width=\linewidth]{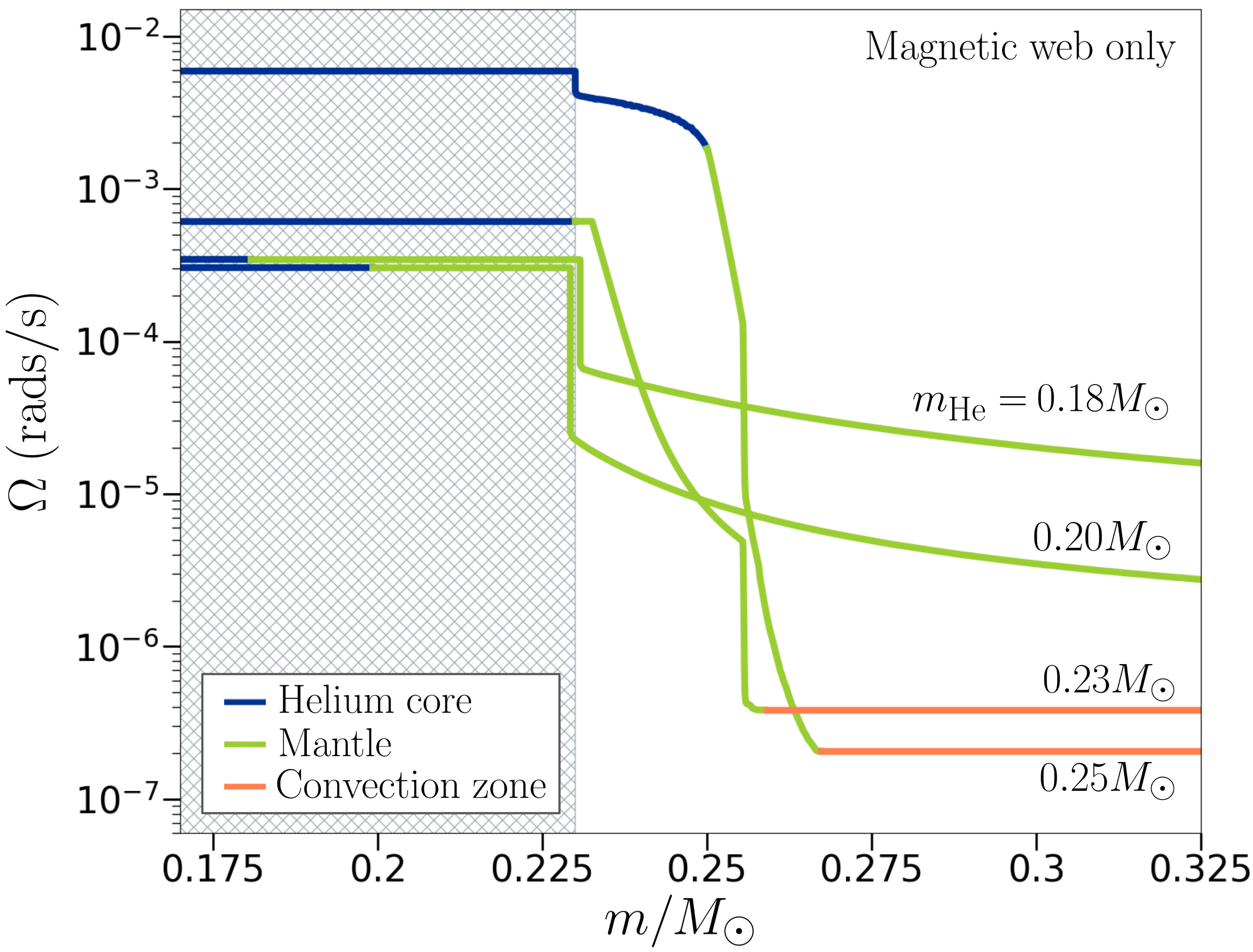}
    \caption{
    A $1.6M_\odot$ stellar model with a magnetic web spanning the mass shells $0\leq m\leq \mt=0.23M_\odot$ (hatched) and with AM transport disabled in radiative regions outside the web.
    Top: Kippenhahn diagram including the MS and RGB phases. 
    Bottom: Rotation profiles $\Omega(m)$ at different evolutionary points on the RGB, with the growing helium core mass $\mHe$ indicated next to each profile.
    }
    \label{fig:MESA_models}
\end{figure}

We use MESA \citep{paxton2010modules,paxton2013modules,paxton2015modules,paxton2018modules,paxton2019modules,jermyn2023modules} to examine the rotational evolution of a star containing a magnetic web. We focus on a $1.6M_\odot$ star representative of the red giant KIC 11515377 which has been asteroseismically inferred to contain $\overline{B}_R\sim 10^5\;$G fields at the edge of its helium core of current mass $\mHe\approx 0.2M_\odot$ \citep{li2022magnetic}. The origin of these fields is consistent with deposition by a core convection zone that extended out to a mass shell $m\approx 0.23M_\odot$ during the MS \citep{li2022magnetic}, which we interpret as evidence for a surviving magnetic web since $\overline{B}_R\gg \overline{B}_{\rm web}\sim30\;$G (\Eq~\ref{eq:Bweb1}).

The star is initialized with uniform rotation rate $\Omega\approx 5\cdot10^{-5}\;\rm{rads/s}$ ($\sim1.5$\,d period) and solar metallicity $Z=0.02$. At the boundaries of convective zones we use exponential convective overshoot with $e$-folding scale $f_{\rm ov}H_{\rm p}$, where $H_{\rm p}$ is the pressure scale-height and $f_{\rm ov}$ is the overshoot parameter (e.g. \citealt{anders2023convective}). We adopt a relatively low $f_{\rm ov}=0.005$ which helps the model fit rotation data for red giants; below we also show a model with $f_{\rm ov}=0.01$.

The magnetic web is implemented as a large AM diffusivity $\nu_{\rm web}=H_{\rm p} v_{\rm A}$ (where $v_{\rm A}=\overline{B}_{R}/\sqrt{4\pi\rho}$) between web boundaries $0\leq m\leq \mt$ with $\mt=0.23M_\odot$. The field strength $\overline{B}_{R}=\sqrt{8\pi P/\beta}$ is prescribed to approximate flux-freezing from the core convective phase, where $P$ is the plasma pressure and $\beta\sim10^{5}$ is the ratio of the plasma pressure to the magnetic pressure. The resulting large $\nu_{\rm web}$ effectively enforces solid body rotation of the region covered by the web.

We first examine an idealized model with AM transport disabled in the radiative zone outside the web (Figure~\ref{fig:MESA_models}). During the early RGB, the web extends into the mantle ($\mt>\mHe$) and enforces corotation of shells $m<\mt$ by redistributing the AM of the contracting core toward the web's outer edge. Since the mass shells covered by the web conserve their total AM $\Jweb$, their rotation rate $\Omega_{\rm web}(t)=\Jweb/\Iweb(t)$ is determined by their evolving moment of inertia $\Iweb(t)$. The magnetic web essentially appends the large moment of inertia of the lower mantle to that of the contracting core. This reduces the spin up of the core by the factor\footnote{Here we approximated $m\propto R^{3}$ in the central region of the MS star, before its contraction into the core.}
\begin{equation}
\label{eq:reductionFactor}
    \chi\equiv\frac{J_{\rm c}}{\Jweb}\frac{\Iweb}{I_{\rm c}}\sim\left(\frac{m_{\rm He}}{\mt}\right)^{5/3}\left[1+\frac{\mt-m_{\rm He}}{m_{\rm He}}\left(\frac{\Rweb}{R_{\rm He}}\right)^2\right]    
\end{equation} 
compared to if the core conserved its initial AM $J_{\rm c}$. The steep drop in density outside the core at $m\gtrsim\mHe$ means that the outer mass shells of the web $\mHe<m<\mt$ occupy a much larger volume and have a much larger lever arm than those inside the core, $\Rweb\gg R_{\rm He}$, and so they dominate $\Iweb$.

\begin{figure}
    \centering
    \includegraphics[width=\linewidth]{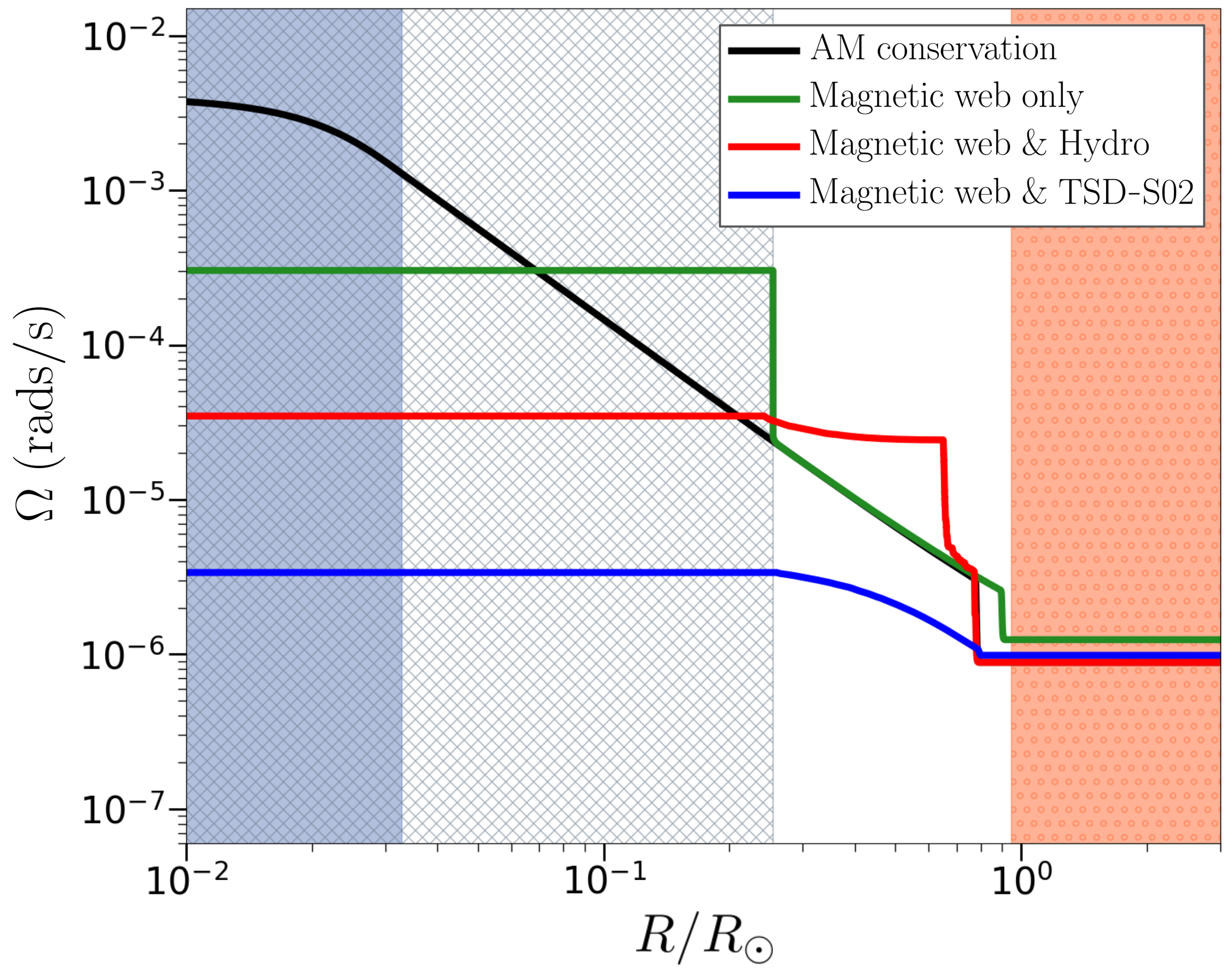}
    \caption{
    Rotation profiles $\Omega(R)$ of the $1.6M_\odot$ star, calculated with different prescriptions for AM transport (see text). Profiles are taken when $\mHe=0.2M_\odot$, representative of the evolutionary point of KIC 11515377. Shaded and hatched regions are the same as in Figure~\ref{fig:MESA_models}. 
    }
    \label{fig:Model_comparison}
\end{figure}

The idealized model is, however, incomplete because it implies a jump in $\Omega$ at $\mt$, i.e. a huge shear at the outer boundary of the web. Turbulent processes likely moderate this shear and extract AM from the web zone, reducing $\Jweb$. To examine this effect, we compare the idealized stellar model with models that include different prescriptions for the turbulent viscosity $\nu_{\rm turb}$ in the radiative zone outside $\mt$. Figure~\ref{fig:Model_comparison} shows the rotation profiles $\Omega(R)$ when $\mHe=0.2$ (the evolutionary point of KIC 11515377) for models that prescribe 
(1) AM conservation for each mass shell (no transport), 
(2) transport solely by the magnetic web, or
(3) transport by both the web and a turbulent viscosity for two versions of $\nu_{\rm turb}$. 
The efficiency of AM extraction from the core can be quantified by the ratio of rotation rates in the core and the envelope, $\Omega_{\rm c}$ and $\Omega_{\rm e}$. The model without any transport gives $\Omega_{\rm c}/\Omega_{\rm e}\sim 3000$. The model with only a magnetic web reduces the ratio to $\Omega_{\rm c}/\Omega_{\rm e}\sim 300$, i.e. $\chi\sim 10$ as anticipated from \Eq~(\ref{eq:reductionFactor}). This is a remarkable reduction in view of the small $(\mt-m_{\rm He})/\mt=0.15$ (the web is nearly buried in the growing core). When turbulent transport is included outside the web, the ratio $\Omega_{\rm c}/\Omega_{\rm e}$ decreases further. We find that including $\nu_{\rm turb}$ based on hydrodynamical mixing processes  \citep{heger2000presupernova,paxton2013modules} gives $\Omega_{\rm c}/\Omega_{\rm e}\sim 30$, and including  $\nu_{\rm turb}$ based on the Tayler-Spruit dynamo \citep{spruit2002dynamo} gives $\Omega_{\rm c}/\Omega_{\rm e}\sim 3$.

\begin{figure}
    \centering
    \includegraphics[width=\linewidth]{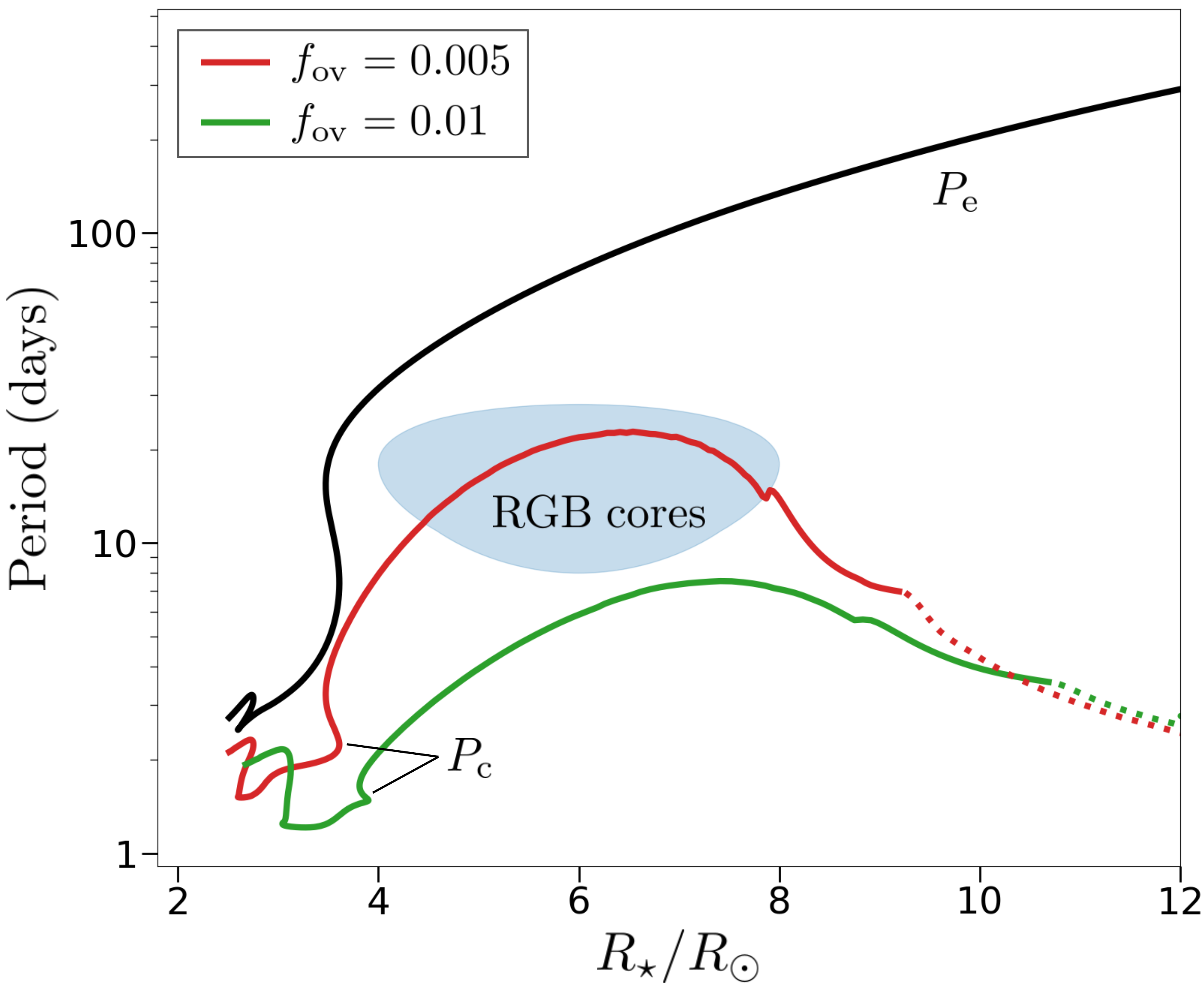}
    \caption{
    Evolution of the rotation period of the envelope $P_{\rm e}$ (black curve) and the core $P_{\rm c}$ (colored curves) in a $1.6M_\odot$ star. On the horizontal axis, we use the growing radius of the star $R_\star$ as a proxy for its age. The stellar model includes AM transport by the magnetic web and turbulent transport outside the web (Tayler-Spruit dynamo
    model). Two evolution histories (red and green) are calculated assuming different values for the convective overshoot parameter $f_{\rm ov}$. The dotted portions of the curves indicate the late evolutionary phase when the web is completely buried in the growing helium core below the hydrogen burning shell. The models are compared with asteroseismic data for $P_{\rm c}$ \citep{mosser2012spin,gehan2018core}, indicated by the blue shaded region.
    }
    \label{fig:DataComparison}
\end{figure}

The model with the greatest AM extraction from the core is consistent with asteroseismic observations of red giant cores \citep{mosser2012spin,gehan2018core}. This is illustrated in Figure~\ref{fig:DataComparison}, which shows the evolution of the rotational period of the core $P_{\rm c}=2\pi/\Omega_{\rm c}$. However, we note that the model predictions are fairly sensitive to the convective overshoot parameter $f_{\rm ov}$. Raising its value to $f_{\rm ov}=0.01$ reduces $P_{\rm c}$ by a factor of $\sim3$. Larger overshoot increases the size of the helium core accumulated from hydrogen burning on the MS but does not affect the mass enclosed by the web $\mt$.\footnote{
Core convection occupies the largest region  $0<m<\mt$ early on the MS when the star briefly burns $^3$He and $^{12}$C out of equilibrium. The value of $\mt$ set at this early stage is insensitive to convective overshoot \citep{li2022magnetic}.} 
This leads to a less dramatic reduction factor $\chi$ due to web coupling (\Eq~\ref{eq:reductionFactor}), so the core spins faster.

As the star keeps evolving on the RGB, the web eventually becomes buried inside the growing helium core ($\mHe$ exceeds $\mt$) and the core-mantle coupling weakens as it then relies on turbulent transport alone. As a result, the model in Figure~\ref{fig:DataComparison} becomes similar to MESA models that prescribe only the Tayler-Spruit dynamo \citep{spruit2002dynamo}, which fail to reproduce the spindown of cores to periods $\sim 100$\,d reported for red clump stars \citep{cantiello2014angular}. Extensions of the model that may explain the late evolution are discussed below.

\section{Discussion}
\label{sec:Discussion}

Magnetic webs are consistent with strong magnetic fields inferred at the boundaries of red giant cores from asteroseismology. We find that magnetic webs also naturally explain efficient AM transport out of stellar cores until they are buried inside the growing core. Our example $1.6M_\odot$ stellar model (Section~\ref{sec:StellarModel}) demonstrates how a single magnetic web (deposited by the core convection on the MS) moderates the rotation of the helium core by coupling it to the lower part of the surrounding radiative mantle, which has a much larger moment of inertia. Turbulent transport operating outside the web further reduces the core rotation to rates comparable with asteroseismic observations during the RGB. 

We expect that a similar evolution broadly occurs for low-mass stars with initial masses above the Kraft break $M\gtrsim 1.3M_\odot$ \citep{kraft1967studies,van2013fast,cantiello2016asteroseismic}. Stars with larger $M$ have larger zones of core convection on the MS, which should leave magnetic webs covering larger regions $0<m<\mt$. Then, the web burial should occur later on the RGB. This trend is consistent with the observed mass dependence of the suppression of dipole-mode oscillations in red giants due to the magnetic greenhouse effect, which is associated with the presence of magnetic fields at $m\approx\mHe$ \citep{fuller2015asteroseismology,stello2016prevalence}.  

Once a web is buried inside a growing helium core, it can no longer assist in extracting AM from the core. Therefore, the single web model is insufficient to explain efficient AM loss at late evolutionary stages, as inferred from slow rotation in red clump stars \citep{mosser2012spin,deheuvels2015seismic} and stellar remnants (white dwarfs, see \citealt{kawaler2014rotation,hermes2017white}). If no webs remain outside the core, boosted turbulent transport mechanisms may need to be invoked \citep{fuller2019slowing}.

However, an alternative possibility is that stars host multiple webs, each providing core-mantle coupling at different evolutionary stages. If every convective phase leaves a magnetic web, the $1.6M_\odot$ red giant would have more webs than assumed in the ``minimal'' model in Section~\ref{sec:StellarModel}. In particular, the fully-convective pre-MS phase may leave behind a global fossil field. The lower portion of this web must be destroyed by the convective core during the MS, and later replaced with a new web deposited by the receding core convection. This would leave the ``core" and ``fossil" webs separated by a narrow gap where differential rotation would occur. These webs would moderate core rotation throughout most of the RGB. During the later RGB, the convective envelope would deposit a third web after it reaches its innermost mass shell $m_{\rm dredge}$ (the first dredge up) and begins to recede upward. This web would recouple the outer helium core to the mantle once $m_{\rm He}>m_{\rm dredge}$. 

The observational evidence for magnetic webs and mild requirements for their formation suggests that webs may be a common feature of radiative zones. This leads to the general picture of a star's radiative zones mostly covered by webs that mark the locations of extinguished convective zones from earlier evolutionary phases. The webs are then separated by gaps with concentrated differential rotation, which keeps neighboring webs disconnected and simultaneously induces turbulent transport in the gaps, facilitating AM exchange between webs.

Our proposed picture for AM transport in stellar radiative zones relies on convective zones leaving behind magnetic fields. However, the process of field deposition by a receding convective boundary is still an open problem \citep{kissin2015spin,braithwaite2017magnetic}. The deposition efficiency likely depends on several factors, including rotation, which can affect the convective dynamo, and the direction of the recession (inward or outward), which is important if magnetic buoyancy plays a significant role. We note that a stable magnetic configuration in the Sun ---a solar magnetic web--- has been long invoked to explain the solid-body rotation of the solar radiative zone (e.g. \citealt{gough1998inevitability}). Its fields must be of fossil origin, i.e. deposited by the pre-MS convection. We also highlight the recent asteroseismic inference of $\overline{B}_R\sim 5\cdot 10^5\;$G in a $\sim 6 M_\odot$ MS star \citep{lecoanet2022asteroseismic}. The field is found in the compositionally stratified layer left behind by the shrinking convective core, which may be interpreted as observational evidence for magnetic web deposition by an actively receding convection zone.

In addition to AM transport, magnetic webs also have broader implications for stellar evolution as they modify mixing processes. Rotational mixing is suppressed inside of webs because shear-driven instabilities are not active, while thermohaline mixing can be substantially modified depending on the local direction of the magnetic field \citep{charbonnel2007inhibition,harrington2019enhanced,fraser2024magnetized}. Furthermore, chemical mixing and AM transport by hydrodynamic waves is also affected since magnetic fields modify the propagation of internal waves \citep{fuller2015asteroseismology,lecoanet2017conversion,rui2023gravity,duguid2024efficient,astoul2025interplay}.  

An important future direction is examining the role of magnetic webs in massive stars. Observations of long-lived surface fields in the radiative envelopes of massive stars suggest that stable magnetic configurations can result from relaxation of fields generated before the MS \citep{braithwaite2017magnetic}.  Magnetic memory of later convection phases requires further study (e.g. \citealt{kissin2018rotation}). Massive stars contain a large core-convection region during the MS, and afterwards host complicated patterns of convective zones. A large web formed after the MS may be broken up by the smaller convective zones, which would leave behind their own, smaller webs. Since nearly all radiative zones will be covered by webs, AM extraction from the core is likely dominated by the conveyor of AM through multiple webs, with turbulent transport operating  between them.

A large uncertainty lies in how the interaction of webs and convection zones leads to the formation, merger, or splitting of webs. Detailed modeling of multiple webs in massive stars could help constrain their rotation and magnetization in the final evolutionary stages. It will have important implications for their binary interactions \citep{sana2012binary}, collapse scenario, and compact remnants (e.g. \citealt{muller2020hydrodynamics,burrows2021core}). Efficient web formation likely leads to magnetized and slowly rotating cores, since rotational decoupling (and destruction of webs) occurs only at the final stages of nuclear burning when evolutionary timescales become short \citep{spruit1998birth,kissin2018rotation}. The efficient loss of core AM in the majority of massive stars may explain the low occurrence of collapsars capable of producing cosmological gamma-ray bursts \citep{macfadyen1999collapsars}. It may also explain why only $\sim 10\%$ of neutron stars are born as magnetars with internal fields $B\sim 10^{16}$\,G \citep{kaspi2017magnetars}, whose formation likely requires fast rotation.

Future work can also help model the evolution of individual webs, which we sketch in Appendix~\ref{ap:webevol}. One uncertainty is how fast web boundaries change in the mass coordinate $m$ due to the buoyant rise of magnetic fields enabled by thermal diffusion \citep{macgregor2003magnetic,braithwaite2008non}. In low mass stars, the web deposited by MS core convection has $\sim 1$\,Gyr to evolve. Its spreading in $m$ would increase the moment of inertia coupled to the helium core and delay web burial. This would lead to even slower core rotation. Another uncertainty concerns AM fluxes across web boundaries, where significant differential rotation occurs. Here, AM can be exchanged by local turbulent transport or filtering of propagating internal gravity waves excited in neighboring convective zones.

\begin{acknowledgments}
We thank Chris Thompson, Selma de Mink, Matteo Cantiello, Daniel Lecoanet, and Jim Fuller for insightful discussions, and Kailey Whitman for help with illustrations. This work is supported by NSF grant AST-2408199. A.M.B. is also supported by NASA grants 21-ATP21-0056, 80NSSC24K1229, and Simons Foundation award No.
446228.

\end{acknowledgments}

%

\vspace{5mm}





\appendix

\section{Response of an axisymmetric magnetic web to changes in rotation rate}
\label{axisym_response}

Consider a star threaded by a stable axisymmetric magnetic field $\overline{\boldsymbol{B}}(R,\theta)$ and initially rotating with a uniform angular velocity $\Omega$ and velocity $\overline{U}_\phi=\Omega R\sin\theta$. Here, we use spherical coordinates $R,\theta,\phi$. Suppose velocity fluctuations $\overline{u}_\phi$ are excited by an axisymmetric torque with density $\tau(R,\theta,t)$. We will parameterize the torque  as 
\beq
  \tau=f(R)\, \frac{\rho\Omega R^2 \sin^2\theta}{\tpump},
\eeq
where $f(R)$ is a dimensionless function and $\rho$ is the fluid density. We will assume that the volume integral of $\tau$ is zero, so the net AM of the fluid remains constant; then, the generated perturbation is similar to differential rotation excited in a star with a contracting core and an expanding envelope. The perturbation of the angular velocity is 
\beq
  \delta\Omega(R,\theta,t)=\frac{\overline{u}_\phi(R,\theta,t)}{ R\sin\theta}.
\eeq
The created non-uniform rotation begins to shear the poloidal magnetic field $\overline\bB_p(R,\theta)$ and induces perturbations $\overline{b}_\phi(R,\theta,t)$ away from the initial equilibrium. Strong stable stratification with a Brunt-V\"ais\"al\"a frequency $N\gg\Omega$ suppresses meridional flows (e.g. thermal winds and circulations) driven by the solenoidal part of the Coriolis force sourced by $\overline{u}_\phi$ \citep{mestel1988mutual,moss1990rotation,charbonneau1993angular}. Then, the equations governing the axisymmetric perturbations are
\begin{align}
\label{eq:dOm}
    \rho R^2 \sin^2 \! \theta\, \partial_t\delta\Omega&=
    \frac{1}{4\pi }\overline{\boldsymbol{B}}_p\cdot \nabla 
    (\overline{b}_\phi R\sin\theta)
    +\tau,\\
\label{eq:dBphi}
    \partial_t  \overline{b}_\phi
    &=R\sin\theta\,\overline{\boldsymbol{B}}_p\cdot \nabla \delta\Omega,
\end{align}
where the diffusive terms are omitted for brevity.

First, consider the simplest problem with a torque $\tau$ switching on suddenly at $t=0$ and remaining steady at $t>0$. It launches a time-dependent perturbation, which eventually relaxes to a static deformation $\bphipert=\Delta\bphipert(R,\theta)$, since any damping eventually suppresses oscillations. This final state has a uniform rotation rate\footnote{More generally, the uniform rotation $\Omega(t)=J/I(t)$ can vary in time if the coupled mass shells change their moment of inertia $I$ while conserving their net AM. Accounting for the radial contraction/expansion of mass shells requires additional terms in \Eq~(\ref{eq:dOm}) and (\ref{eq:dBphi}) \citep{gouhier2022angular}, but will not change the main results presented here.}
($\delta\Omega=0$) equal to the initial $\Omega$ and a static deformation described by a particular solution of Equations~(\ref{eq:dOm}) and (\ref{eq:dBphi}):
\begin{equation}
    \frac{1}{4\pi}\overline{B}_s\frac{\partial}{\partial s} \left[r(s)\, 
    \Delta\bphipert
    \right]+\tau(s)=0, \qquad r\equiv R\sin\theta,
\end{equation}
where coordinate $s$ runs along the poloidal field line. Using $ds/\overline{B}_s=dR/\overline{B}_R$ and $dm=4\pi\rho R^2 dR$, we estimate
\begin{align}
\label{eq:DeltaB_}
    |\Delta \bphipert|\lesssim \frac{\Omega}{R}\int \frac{4\pi\rho R^2|f(R)|}{\overline{B}_R \tpump}  dR\sim \frac{m_{\rm web}\Omega}{\Rweb\overline{B}_R\tpump}.
\end{align}

The full time-dependent solution for $\bphipert(s,t)$ is a sum of the particular solution $\Delta\bphipert(s)$ (with $\delta\Omega=0$) and the homogeneous solution $\delta \bphipert(s,t)$ (with $\tau=0$ and non-zero $\delta\Omega$), which describes propagating torsional Alfv\'en waves,
\beq
  \bphipert=\Delta \bphipert+\delta \bphipert.
\eeq
The wave amplitude $\delta \overline{b}_{\phi,\max}$ on a field line is determined by the difference between the initial condition $\bphipert=0$  and the new equilibrium $\bphipert=\Delta \bphipert$ that balances the torque, so $\delta \overline{b}_{\phi,\max}=|\Delta \bphipert|$.

As toroidal Alfv\'en waves are ducted along poloidal field lines, each poloidal flux surface hosts an oscillating perturbation $\delta \bphipert$ that behaves similarly to the toy model in Section~\ref{toy}. However, there are now many oscillators with different Alfv\'en timescales. As waves on neighboring flux surfaces gradually go out of phase, they develop transverse gradients of $\delta\bphipert$ that eventually grow sufficiently large for any finite magnetic diffusivity $\eta$ (or viscosity $\nu$) to damp the waves \citep{ionson1978resonant,heyvaerts1983coronal,cally1991phase,charbonneau1993angular}. This damping via phase-mixing occurs after timescale
\beq
    t_{\rm ph}=t_{\rm A}\left(\frac{\min\{t_\eta, t_\nu\}}{t_{\rm A}q_{\rm A}^2}\right)^{1/3}
    =100\;\mathrm{yr}\left(\frac{t_{\rm A}}{1\rm yr}\right)^{2/3}\left(\frac{\min\{t_\eta, t_\nu\}}{10^8\;\rm yr}\right)^{1/3}\left(\frac{10}{q_{\rm A}}\right)^{2/3},
\eeq
where $t_\nu=R^2/\nu$, $t_\eta=R^2/\eta$, and $q_{\rm A}$ is a dimensionless measure of the transverse gradients of $\overline{\bB}_{p}$ \citep{spruit1999differential}. This timescale is shorter than the timescale of stellar evolution, so any excited Alfv\'en waves are quickly damped.  

In a real star, the source pumping differential rotation appears gradually (on the stellar evolution timescale) rather than abruptly. Therefore, in a more realistic model, the torque density $\tau$ grows from zero on a long timescale $\tpump$. Then, the evolution of perturbations on each flux surface is similar to the toy model described in Section~\ref{toy}. In particular, the amplitude of excited waves is smaller than the final static deviation $\Delta \overline{b}_{\phi}$ by a factor of $t_{\rm A}/\tpump \ll 1$. Thus, torsional Alfv\'en waves have a negligible amplitude even when their damping via phase mixing is neglected.

In summary, the response of a magnetic web to the pumping of differential rotation in a star with $t_{\rm A}\ll \tpump$ is well described as a quasistatic deformation $\Delta \overline{b}_{\phi}$ from the initial equilibrium. This deformation is given by \Eq~(\ref{eq:DeltaB_}).

\section{Evolution of individual magnetic webs}
\label{ap:webevol}

Large-scale magnetic fields do not experience significant ohmic diffusion, so they are practically frozen in the fluid. Therefore, our stellar model in Section~\ref{sec:StellarModel} assumed that the magnetic web stayed attached to the mass shells where it was deposited. However, a more detailed model can allow a slow drift of the web boundary in the mass coordinate $m$ due to the buoyancy of magnetic fields. In a stably stratified radiative zone, buoyancy is enabled by the thermal/compositional diffusivity $\kappa$ when the stellar layers are thermally/compositionally stratified \citep{macgregor2003magnetic,braithwaite2008non}. The evolution of a web boundary by this process can be written as
\begin{align}
    \frac{dm_i(t)}{dt}=\frac{\partial m}{\partial R}v_{\rm rise}(m_i),\quad \mathrm{i=b,\;t,}
\end{align}
where $v_{\rm rise}$ is the rise speed of the web's magnetic fields at its bottom and top boundaries $\mb$ and $\mt$. Its maximum value is found for an isolated horizontal magnetic flux tube of radius $L$: $v_{\rm rise}\sim \kappa g/\beta N^2L^2$ \citep{macgregor2003magnetic,braithwaite2008non}, where $g$ is the gravitational acceleration.  For typical values in a thermally stratified zone of a low mass star, a web may be displaced by a significant fraction of a pressure scale height $H_{\rm p}$ over a star's age $t$ for the maximum $v_{\rm rise}$,
\begin{align}
    \frac{v_{\rm rise}t}{H_{\rm p}}\sim \frac{\kappa t}{\beta L^2}\sim0.1\left(\frac{\kappa}{10^7\;\rm cm^2s^{-1}}\right)\left(\frac{t}{10^9\;\rm yr}\right)\left(\frac{\beta}{10^{5}}\right)^{-1}\left(\frac{L}{0.1R_\odot}\right)^{-2}.
\end{align}
However, note that rise speeds may be lower because of the effective drag that is induced by the replacement of rising magnetized fluid by sinking unmagnetized fluid \citep{cantiello2011magnetic}.

Web boundaries can also change if part of a magnetic configuration becomes too weak to stop the development of differential rotation, so this part may eventually be destroyed and the web boundary contracts to where the condition $\overline{B}_{R}/\overline{B}_{\rm web}>1$ is still satisfied. This partial loss of the web may happen in expanding regions of an evolving star where the local magnetic field strength decreases as $\overline{B}_R(R)\propto R^{-2}$ due to flux-freezing. Since the minimum radial field for a healthy web decreases slower with radius, $\overline{B}_{\rm web}\propto R^{-3/2}$ (\Eq~\ref{eq:Bweb}), webs in expanding regions become less sturdy (i.e. the ratio $\overline{B}_{R}/\overline{B}_{\rm web}\propto R^{-1/2}$ decreases). 

Accurate models for web evolution may help detailed comparison with observations. In particular, for low mass stars, the web extension radius $R_{\rm web}$ affects the core spin and the timing of the web's burial within the growing core mass $m_{\rm He}$. Thus, tracking changes of a web boundary from its  original mass coordinate is important for a closer comparison with asteroseismic observations of core rotation rates and magnetic field strengths at core boundaries.


\bibliography{refs}{}
\bibliographystyle{aasjournal}



\end{document}